\documentclass
[aps,pra,amsmath,amsfonts,amssymb,twocolumn,superscriptaddress,showpacs]{revtex4}%
\usepackage{amsfonts}
\usepackage{amsmath}
\usepackage{amssymb}
\usepackage{graphicx}%
\providecommand{\U}[1]{\protect\rule{.1in}{.1in}}
\begin{document}
\title{Purity as a witness for initial system-environment correlations\\in open-system dynamics}
\author{D. Z. Rossatto}
\affiliation{Departamento de F\'{\i}sica, Universidade Federal de S\~{a}o Carlos, CEP
13565-905, S\~{a}o Carlos, SP, Brazil}
\author{T. Werlang}
\affiliation{Departamento de F\'{\i}sica, Universidade Federal de S\~{a}o Carlos, CEP
13565-905, S\~{a}o Carlos, SP, Brazil}
\author{L. K. Castelano}
\affiliation{Departamento de F\'{\i}sica, Universidade Federal de S\~{a}o Carlos, CEP
13565-905, S\~{a}o Carlos, SP, Brazil}
\author{C. J. Villas-Boas}
\affiliation{Departamento de F\'{\i}sica, Universidade Federal de S\~{a}o Carlos, CEP
13565-905, S\~{a}o Carlos, SP, Brazil}
\author{F. F. Fanchini}
\affiliation{Departamento de F\'{\i}sica, Universidade Federal de Ouro Preto, CEP
35400-000, Ouro Preto, MG, Brazil}

\begin{abstract}
We study the dynamics of a two-level atom interacting with a Lorentzian
structured reservoir considering initial system-environment correlations. It
is shown that under strong system-reservoir coupling the dynamics of purity
can determine whether there are initial correlations between system and
environment. Moreover, we investigate the interaction of two two-level atoms
with the same reservoir. In this case, we show that besides
determining if there are initial system-environment correlations, the dynamics
of the purity of the atomic system allows the identification of the distinct correlated initial states.
 In addition, the dynamics of quantum and classical correlations
is analyzed.

%We study the dynamics of an two-level atom interacting with a reservoir at zero temperature without the Markov and Born approximations. It is shown that under strong system-reservoir coupling the dynamics of purity can determine whether there are initial correlations between system and environment. Moreover, we consider the interaction of two two-level atoms with the same reservoir and we study the dynamics of quantum and classical correlations and the purity of the atomic system. In this case, we show that besides determining if there are initial system-environment correlations, the dynamics of these quantities allows us to identify the distinct correlated initial states.

\end{abstract}

\maketitle

\section{Introduction}

The preparation of the initial state plays a key role in the implementation of
any experiment. As recently noted, \cite{Roleprep} different procedures of the preparation of
initial states can lead to nontrivial differences in the experimental results
due to interaction with the environment. These differences arise from initial
correlations between the system and its environment. Although
the assumption of uncorrelated system-reservoir initial states is widely used,
it is not always well justified, specially when the system strongly interacts
with the environment \cite{pechukas}. Thus, the influence of initial
correlations on the dynamics of open quantum systems has been intensively
studied
\cite{recent1,recent2,recent4,recent5,eofchina,recent6,QDchina,Linearmaps,notpositimaps}%
. In Ref.~\cite{trl}, the authors verified that the linear transient response
of a two-level system coupled to a reservoir differs significantly if the
initial state is correlated or not. Another important issue concerns about the
possibility of associating a completely positive map to a quantum dynamics,
which is possible only when there are no initial quantum correlations between
system and reservoir \cite{VanishingQD}. Moreover, the effects of the initial
correlations were recently observed experimentally by using an all-optical
apparatus to probe the evolution of the two-qubit polarization entangled state
\cite{correxp}.

The usual way of measuring correlations in quantum systems is by mutual
information \cite{nielsen}. Although it is a well established measure, the
mutual information does not distinguish between quantum and classical aspects
of correlation, which is an important issue of quantum information theory. In
recent years, the study of measures that can differentiate among these two
distinct aspects of correlation has been intensified
\cite{bennet,zurek,varios}, specially after the discovery that mixed
unentangled states can also have non-classical correlation \cite{bennet,zurek}.
The more widespread measure of total quantum correlation in a bipartite
quantum system is the quantum discord, which was proposed by Ollivier and
Zurek \cite{zurek}. A related quantity concerning classical correlations
was also proposed by Henderson and Vedral \cite{henderson}. A possible way to
determine the existence of correlated initial states was suggested by E.-M.
Laine \textit{et al} \cite{witnessbreuer} and it is based on the trace
distance, a measure of distinguishability between two quantum states
\cite{breuer-prl}. They have shown that if the trace distance increases over
its initial value, the system-reservoir state was initially correlated.
However, this approach does not identify the nature of the initial
system-environment correlations, that is, the quantum and classical aspects.
 %{This task is one of our purposes in this work.}

In this work we study the dynamics of a system consisting of a two-level atom
(qubit) interacting with a zero-temperature environment without the Markov and
Born approximations. We show that the system purity dynamics can be used as
a witness for initial correlations between system and environment.
Furthermore, by introducing a second atom (probe qubit) interacting with the
same environment, we are able to determine the different initial
system-environment correlations through the dynamics of the purity between the
qubit and the probe qubit. We also study the dynamics of entropy and classical
and quantum correlations of the atomic system. In this case, we show that
these dynamics can be useful to identify and to classify the initial
correlations between system and environment.

\section{Theoretical Model}

In this section, we first describe theoretically some important quantities
that are employed to quantify quantum and classical correlations within
quantum systems. These quantities are useful in our work because they
exhibit signatures of initial system-environment correlations, as demonstrated by our
results. Moreover, we present the model used to describe two atoms coupled to a common environment at zero
temperature. The approach to solve the dynamics of the system-environment is
discussed as well.

\subsection{Quantum and Classical Correlations}

\textit{Quantum Discord:} Quantum correlations have played an important role
in quantum information and communication theory \cite{nielsen}. If we consider
for instance a bipartite quantum state described by the density matrix
$\rho_{AB}$, the total correlations between the subsystems $A$ and $B$ can be
computed by using the mutual information \cite{nielsen}:
\begin{equation}
\mathcal{I}\left(  \rho_{AB}\right)  =\mathcal{S}\left(  \rho_{A}\right)
+\mathcal{S}\left(  \rho_{B}\right)  -\mathcal{S}\left(  \rho_{AB}\right)  ,
\label{IM}%
\end{equation}
where $\rho_{i}=\mbox{Tr}_{j}\left(  \rho_{ij}\right)  $ and $\mathcal{S}%
(\rho)=-\mbox{Tr}\left(  \rho\log_{2}\rho\right)  $ is the von-Neumann
entropy. A measure of classical correlations present in the quantum state
$\rho_{AB}$ is defined as \cite{henderson}:
\begin{equation}
\mathcal{J}\left(  \rho_{AB}\right)  =\mathcal{S}\left(  \rho_{A}\right)
-\mathcal{S}\left(  \rho_{A|B}\right)  , \label{CC}%
\end{equation}
where $\mathcal{S}\left(  \rho_{A|B}\right)  \equiv\min_{\left\{
M_{b}\right\}  }\sum_{b}p_{b}\mathcal{S}\left(  \rho_{A|b}\right)  $ is the
quantum conditional entropy. The minimization is given over generalized
measurements $\left\{  M_{b}\right\}  $ \cite{nielsen}, with $\sum_{b}%
M_{b}=\boldsymbol{1}_{B}$, $M_{b}\geq0$ for all $b$, and
\[
\rho_{A|b}=\frac{(\boldsymbol{1}_{A}\otimes M_{b})\rho_{AB}(\boldsymbol{1}%
_{A}\otimes M_{b})}{\mbox{Tr}\left[  (\boldsymbol{1}_{A}\otimes M_{b}%
)\rho_{AB}(\boldsymbol{1}_{A}\otimes M_{b})\right]  },\label{rhomedido}%
\]
is the reduced density operator of $A$ after obtaining the outcome $b$ in $B$.
For two-qubit systems, the minimization over generalized measurements is
equivalent to a minimization over projective measurements \cite{Hamieh}. In
this work, we use numerical minimization to compute the classical correlations.

Using the results above, the quantum discord \cite{zurek} is defined as
\begin{equation}
D\left(  \rho_{AB}\right)  \equiv\mathcal{I}\left(  \rho_{AB}\right)
-\mathcal{J}\left(  \rho_{AB}\right)  , \label{QD}%
\end{equation}
and it assumes equal values, irrespective of whether the measurement is
performed on the subsystem $A$ or $B$, only when $\mathcal{S}\left(  \rho
_{A}\right)  =\mathcal{S}\left(  \rho_{B}\right)  $ \cite{henderson}\newline

\textit{Entanglement:} A special class of quantum correlated states are the
entangled states. A quantum bipartite state $\rho_{AB}$ is said entangled if and
only if it cannot be written as a separable state $\rho_{AB}=\sum_{j}p_{j}%
\rho_{j}^{A}\otimes\rho_{j}^{B}$, where $\sum_{j}p_{j}=1$ \cite{entangle}. In
this work we use the entanglement of formation (EoF) \cite{Bennett} as a
measure of entanglement, which is defined as $EoF(\rho_{AB})=\min\sum_{j}%
p_{j}\mathcal{S}(\left\vert \psi_{j}\right\rangle )$, where the minimization
is over all ensembles of pure states $\left\{  p_{j},\left\vert \psi
_{j}\right\rangle \right\}  $ such that $\rho_{AB}=\sum_{j}p_{j}\left\vert
\psi_{j}\right\rangle \left\langle \psi_{j}\right\vert $. For a pair of qubits,
the EoF is a monotonically increasing function of the concurrence $C$
\cite{Woo98}
\begin{equation}
EoF\left(  \rho_{AB}\right)  =-f(C)\log_{2}f(C)-(1-f(C))\log_{2}(1-f(C)),
\label{EoF}%
\end{equation}
where $f(C)=\left(  1+\sqrt{1-C^{2}}\right)  /2$ and $C=\max\left\{
0,\lambda_{1}-\lambda_{2}-\lambda_{3}-\lambda_{4}\right\}  $ with $\lambda
_{1}$, $\lambda_{2}$, $\lambda_{3}$ and $\lambda_{4}$ the square roots of the
eigenvalues, in decreasing order, of the matrix $R=\rho\left(  \sigma_{y}%
^{1}\otimes\sigma_{y}^{2}\right)  \rho^{\ast}\left(  \sigma_{y}^{1}%
\otimes\sigma_{y}^{2}\right)  $. Here $\rho^{\ast}$ denotes the complex
conjugation of the matrix $\rho$ in the computational basis $\left\{
\left\vert 00\right\rangle ,\left\vert 01\right\rangle ,\left\vert
10\right\rangle ,\left\vert 11\right\rangle \right\}  $. While all pure
quantum correlated states are entangled states, there are quantum mixed states
with null entanglement and a positive value of quantum discord. This result exemplifies
that not all quantum correlations can be described by entanglement.

\subsection{Dissipative dynamics of two atoms coupled to a common environment}

In this subsection, we describe the dynamics of two non-interacting two-level
atoms coupled to a zero-temperature common environment - the damped
Jaynes-Cummings model. The environment is represented by a bath of harmonic
oscillators with the Hamiltonian given by ($\hbar=1$) $H_{R}=\sum_{k}%
\omega_{k}a_{k}^{\dagger}a_{k}$, where $a_{k}^{\dagger}$ ($a_{k}$) is the
creation (annihilation) operator for the field mode $k$ with frequency
$\omega_{k}$. Our system consists of two two-level atoms whose Hamiltonian (in
computational basis) is $H_{0}=\omega_{0}\left(  \sigma_{+}^{1}\sigma_{-}%
^{1}+\sigma_{+}^{2}\sigma_{-}^{2}\right)  $, where $\omega_{0}$ is the Bohr
frequency and $\sigma_{+}^{j}$ and $\sigma_{-}^{j}$ are the
Pauli raising and lowering operators of the $j$-atom, respectively. The Hamiltonian of the
system plus environment is $H=H_{S}+H_{R}+H_{SR}$ with
\begin{equation}
H_{SR}=\left(  \sigma_{+}^{1}+\sigma_{+}^{2}\right)  \sum_{k}g_{k}a_{k}+h.c.,
\label{HSR}%
\end{equation}
where $g_{k}$ is the coupling constant.

The dynamics can be solved using auxiliary variables defined from the
properties of the spectral distribution through the pseudo-mode approach
\cite{pseudom}. Here we assume that the two atoms interact resonantly with a
Lorentzian structured reservoir, resulting in the following pseudo-mode master
equation in the interaction picture \cite{pseudom}
\begin{equation}
\frac{d\rho}{dt}=-i\left[  V,\rho\right]  +\frac{\Gamma}{2}\left(  2a\rho
a^{\dagger}-a^{\dagger}a\rho-\rho a^{\dagger}a\right)  , \label{drhodt}%
\end{equation}
where $\rho$ is the density operator for the two atoms and the pseudo-mode,
$a$ ($a^{\dagger})$ is annihilation (creation) operator of the pseudo-mode and
$V$ is the interaction Hamiltonian of the atomic system and the pseudo-mode.
The spectral distribution associated with the pseudo-mode is
\begin{equation}
J(\omega)=\frac{\Omega^{2}}{\pi}\frac{\Gamma}{(\omega-\omega_{0})^{2}%
+\Gamma^{2}/4}, \label{spectral}%
\end{equation}
where $\Gamma$ is the pseudo-mode decay rate and $\Omega$ is the coupling
constant between the pseudo-mode and the two atoms. In this model, there are two regimes to be considered \cite{pseudom}: \textit{(i)} a strong-coupling regime that
occurs when $\Gamma<2\Omega$, and \textit{(ii)} a weak-coupling regime for
$\Gamma>2\Omega$. As discussed in Ref.~\cite{pseudom}, for strong-coupling
regime, the atomic dynamics presents non-Markovian features and when the
coupling becomes weak, the Markovian behavior is recovered. In all cases
studied in this work, we assume that the total system has at most one
excitation, so that the Hamiltonian $V$ can be written as \cite{pseudom}
\begin{equation}
V=\Omega\left(  \sigma_{+}^{1}+\sigma_{+}^{2}\right)  a+h.c. \label{Vint}%
\end{equation}
This approach does not rely on either the Born or the Markov approximation and
therefore it is possible to solve the dynamics without performing any further approximation.

\section{Results and Discussion}

One important issue in the theory of open quantum systems is the ability of
determining the origin of the decoherence of a qubit \cite{breuerb}. For
instance, if we start with the single qubit mixed state
\begin{equation}
\rho_{S}\left(  0\right)  =\alpha^{2}\left\vert g\right\rangle \left\langle
g\right\vert +\left(  1-\alpha^{2}\right)  \left\vert e\right\rangle
\left\langle e\right\vert , \label{rhos}%
\end{equation}
where $\alpha\in\left(  0,1\right) $, we are not able to determine how this state was initially prepared.
For example, the qubit state described by Eq.~(\ref{rhos}) can
be obtained from different situations:

a) the state was prepared without any correlation with the environment
\begin{equation}
\rho_{a}=\rho_{S}\left(  0\right)  \otimes\left\vert 0\right\rangle
_{E}\left\langle 0\right\vert , \label{rhoa}%
\end{equation}
where $\left\vert n\right\rangle _{E}$ means the state of the environment with
$n$ excitations;

b) the system was classically correlated with the environment
\begin{equation}
\rho_{b}=\alpha^{2}\left\vert g\right\rangle \left\langle g\right\vert
\otimes\left\vert 1\right\rangle _{E}\left\langle 1\right\vert +\left(
1-\alpha^{2}\right)  \left\vert e\right\rangle \left\langle e\right\vert
\otimes\left\vert 0\right\rangle _{E}\left\langle 0\right\vert ; \label{rhob}%
\end{equation}

c) the system had quantum correlations and null entanglement with the
environment
\begin{equation}
\rho_{c}=\alpha^{2}\left\vert g\right\rangle \left\langle g\right\vert
\otimes\left\vert \phi\right\rangle _{E}\left\langle \phi\right\vert +\left(
1-\alpha^{2}\right)  \left\vert e\right\rangle \left\langle e\right\vert
\otimes\left\vert 0\right\rangle _{E}\left\langle 0\right\vert , \label{rhoc}%
\end{equation}
with $\left\vert \phi\right\rangle _{E}=\left(  \left\vert 0\right\rangle
_{E}+\left\vert 1\right\rangle _{E}\right)  /\sqrt{2}$;

d) the system had non-null entanglement with the environment
\begin{equation}
\rho_{d}\left(  0\right)  =\left\vert \Psi\left(  0\right)  \right\rangle
\left\langle \Psi\left(  0\right)  \right\vert , \label{rhod}%
\end{equation}
being $\left\vert \Psi\left(  0\right)  \right\rangle =\alpha\left\vert
g\right\rangle \left\vert 1\right\rangle _{E}+\sqrt{1-\alpha^{2}}\left\vert
e\right\rangle \left\vert 0\right\rangle _{E}$.

In all states above, tracing over the environment variables, we end up
with the same density operator given in Eq.~(\ref{rhos}). Then, as the
information about the environment is not directly accessible, any measure
carried out on the qubit state can not distinguish whether the system was
initially correlated to the environment or not. According to
Ref.~\cite{witnessbreuer}, the dynamics of the qubit can be used to identify
the presence of initial correlations among system and environment. However,
there is only the possibility of determining if the system was initially correlated to the
environment or not.
%, being not able to distinguish the different aspects of the initial correlations.
Based on this approach, we might say if the qubit initial state
described by Eq.~(\ref{rhos}) was prepared in that way or it was a result from the
interaction with the environment.

In this work, we employ the dynamics of the system
purity to witness the system-environment initial correlation.
  The purity $P$ of a quantum state $\rho$ is defined as $P\left(
\rho\right)  =\mbox{Tr}\left(  \rho^{2}\right)  $ and a quantum state is pure
if and only if the density operator $\rho$ is idempotent, that is, $\rho
^{2}=\rho$.

\begin{figure}
[th]
\begin{center}
\includegraphics[
height=3.3486in,
width=2.4483in
]%
{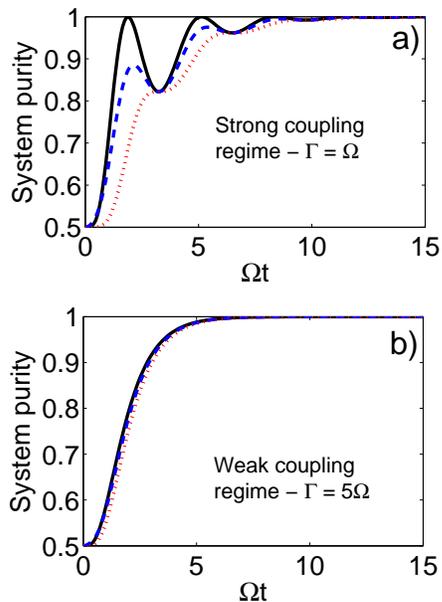}%
\caption{(color online) Time evolution of the one-qubit purity for (a)
strong-coupling $\Gamma=\Omega$ and (b) weak-coupling $\Gamma=5\Omega$
considering the initial states $\rho_{a}$ (solid line), $\rho_{b}=\rho_{d}$
(dotted line) and $\rho_{c}$ (dashed line). We fixed $\alpha^{2}=0.5$.}%
\label{systempur}%
\end{center}
\end{figure}

We begin our analysis by studying the dynamic behavior of the purity $P$ of a
qubit due to the interaction with the environment. We numerically calculate
the dynamics of $P$ for the initial states given in Eqs.~(\ref{rhoa}%
)-(\ref{rhod}), for both strong and weak coupling regimes. In
Fig.~\ref{systempur}, one can notice that the behavior of the dynamics of the
system purity for the uncorrelated state is different from the correlated
ones. However, the dynamics of the entangled initial state $\rho_{d}$
coincides with the classically correlated state $\rho_{b}$.
These states differ from each other by two extra terms that appear in $\rho_{d}$.
For our type of interaction, these extra terms evolve in time not affecting the dynamics of the reduced system,
where the environment has been traced out. In this situation, the dynamics of the system purity for both $\rho_{b}$ and $\rho_{d}$ becomes indistinguishable.
Based on our results, we see that the
dynamics of the system purity can be used as a measure to determine the
system-environment initial correlation, however the kind of initial system-reservoir correlation is not well determined.
Moreover, we observe in Fig.~\ref{systempur}(b) that the curves become
indistinguishable when the system-environment coupling is weak, that is, the
initial correlation does not interfere too much in the system dynamics, so the
supposition of a non-correlated initial state is well justified in this case
\cite{pechukas}. In addition, the purity asymptotically assumes the same value
($P=1$), regardless the initial state. This is expected since we are assuming
a zero-temperature reservoir and thus the asymptotic state of the qubit is the
ground state $\left\vert g\right\rangle $. Although the Fig.~\ref{systempur}
refers to a particular value of the parameter $\alpha=0.5$, our results were
checked for different values of $\alpha$ between $0$ and $1$ and the same
characteristics were observed.%

%TCIMACRO{\FRAME{fthFU}{2.4483in}{3.3486in}{0pt}{\Qcb{(color online) Time
%evolution of the one-qubit purity for (a) strong-coupling $\Gamma=\Omega$ and
%(b) weak-coupling $\Gamma=5\Omega$ considering the initial states $\rho_{a}$
%(solid line), $\rho_{b}=\rho_{d}$ (dotted line) and $\rho_{c}$ (dashed line).
%We fixed $\alpha^{2}=0.5$.}}{\Qlb{systempur}}{system_purity.eps}%
%width 2.4483in;  height 3.3486in;  depth 0pt;  original-width 4.8395in;
%original-height 6.6409in;  cropleft "0";  croptop "1";  cropright "1";
%cropbottom "0";  filename 'system_purity.eps';file-properties "XNPEU";}} }%
%BeginExpansion
%EndExpansion

\begin{figure}
[th]
\begin{center}
\includegraphics[
height=3.3477in,
width=2.2943in
]%
{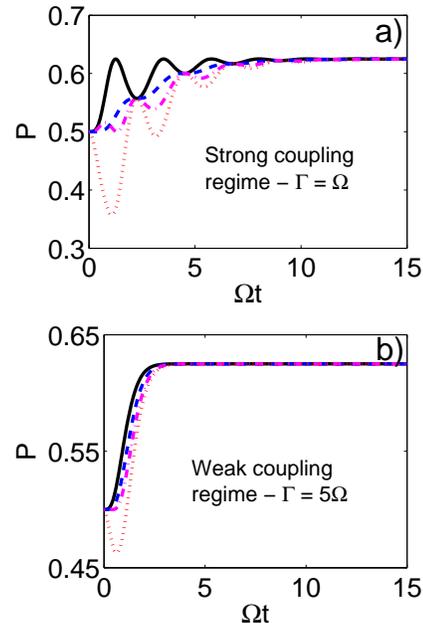}%
\caption{(color online) Time evolution of the two-qubit purity $P$ for (a)
strong-coupling $\Gamma=\Omega$ and (b) weak-coupling $\Gamma=5\Omega$
considering the initial states $\rho_{a}$ (solid line), $\rho_{b}$ (dotted
line), $\rho_{c}$ (dashed line) and $\rho_{d}$ (dash-dot line). We fixed
$\alpha^{2}=0.5$.}%
\label{Fig2}%
\end{center}
\end{figure}

\begin{figure}
[th]
\begin{center}
\includegraphics[
height=2.3419in,
width=3.3287in
]%
{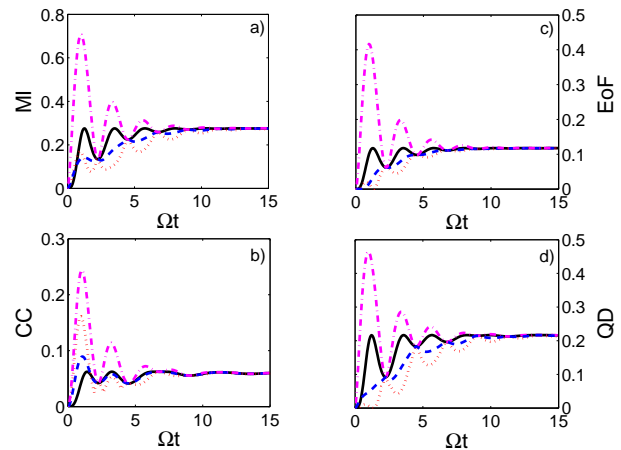}%
\caption{(color online) Dynamics of the (a) mutual information, (b) classical
correlation, (c) entanglement of formation, (d) quantum discord for two-qubits
and for $\rho_{a}$ (solid line), $\rho_{b}$ (dotted line), $\rho_{c}$ (dashed
line) and $\rho_{d}$ (dash-dot line) with $\alpha^{2}=0.5$ e $\Gamma=\Omega$.}%
\label{Fig3}%
\end{center}
\end{figure}

By employing a second qubit (probe) interacting with the same environment and
considering the initial states as $\rho=\left\vert g\right\rangle \left\langle
g\right\vert \otimes\rho_{i}$ $\left(  i=a,b,c,d\right)  $, we numerically evaluate the dynamics of the purity for the two qubits as shown in
Fig.~\ref{Fig2}. For this situation, the dynamics of $P$ have different behavior for
each initial state. Therefore, we are able to determine the initial
correlation between system and environment by observing the system purity dynamics.
To understand this phenomenon, we need to analyze the role of a common environment.
Although both qubits do not interact with each other directly, they are correlated because both qubits interact with the same
reservoir. Such behavior has been observed in different physical systems.\cite{commonb}
Moreover, the common reservoir produces correlations between the qubits and
these correlations depend on the initial state of the system-reservoir.
Such a dependency is more evident when the interaction between the qubits
and the reservoir becomes stronger. Thus, the stronger the interaction, the
better the distinction among the dynamics for each initial state. Besides, the
initial correlation of the original qubit and the environment can be
dynamically transferred to the second qubit in such a way that the purity of
both qubits can produce a signature of the kind of initial correlation among
the first qubit and the environment.
In other words, the two extra terms of the initial state $\rho_{d}$ induce correlations that are captured by the whole system (two atoms) purity.
Again, in the asymptotic regime, the
values of these quantities do not depend on the initial state.
In Fig.~\ref{Fig3}, we plot the (a) mutual information, (b) classical
correlation, (c) entanglement of formation, and (d) quantum discord for
two-qubits as a function of dimensionless time $\Omega t$ for the same initial
states of Figs.~\ref{systempur} and \ref{Fig2}. As already seen in the purity
dynamics, each initial state leads to a particular dynamic behavior of the
correlation measurements. Another interesting aspect about these quantities is
their usefulness to improve the identification of a
particular state. For instance, the dynamics of the purity for the states
$\rho_{c}$ and $\rho_{d}$ are very similar (see Fig.~\ref{Fig2}), which makes
difficult the determination of the initial correlated state. On the other
hand, the dynamics of the other quantities shown in Fig.~\ref{Fig3} for the
same states $\rho_{c}$ and $\rho_{d}$ are very distinct, thereby improving the
ability of specifying the initial correlated state.%

%TCIMACRO{\FRAME{fthFU}{2.2943in}{3.3477in}{0pt}{\Qcb{(color online) Time
%evolution of the two-qubit purity $P$ for (a) strong-coupling $\Gamma=\Omega$
%and (b) weak-coupling $\Gamma=5\Omega$ considering the initial states
%$\rho_{a}$ (solid line), $\rho_{b}$ (dotted line), $\rho_{c}$ (dashed line)
%and $\rho_{d}$ (dash-dot line). We fixed $\alpha^{2}=0.5$.}}{\Qlb{Fig2}%
%}{fig2.eps}{\special{ language "Scientific Word";  type "GRAPHIC";
%width 2.2943in;  height 3.3477in;  depth 0pt;  original-width 4.5333in;
%original-height 6.6409in;  cropleft "0";  croptop "1";  cropright "1";
%cropbottom "0";  filename 'fig2.eps';file-properties "XNPEU";}} }%
%BeginExpansion
%EndExpansion

%

%TCIMACRO{\FRAME{fthFU}{3.3287in}{2.3419in}{0pt}{\Qcb{(color online) Dynamics
%of the (a) mutual information, (b) classical correlation, (c) entanglement of
%formation, (d) quantum discord for two-qubits and for $\rho_{a}$ (solid line),
%$\rho_{b}$ (dotted line), $\rho_{c}$ (dashed line) and $\rho_{d}$ (dash-dot
%line) with $\alpha^{2}=0.5$ e $\Gamma=\Omega$.}}{\Qlb{Fig3}}{fig3.eps}%
%width 3.3287in;  height 2.3419in;  depth 0pt;  original-width 14.2227in;
%original-height 6.7741in;  cropleft "0";  croptop "1";  cropright "1";
%cropbottom "0";  filename 'fig3.eps';file-properties "XNPEU";}} }%
%BeginExpansion
%EndExpansion

To determine if there is or not initial correlation between the qubit and its
environment, it would be enough to compare the dynamics of the qubit purity
obtained experimentally with the theoretical result. If it is necessary to
obtain a more precise information about the type of the initial correlation,
we must employ a probe qubit interacting with the same environment and then
compare the dynamics of one of those physical quantities discussed above with
the theoretical results. In order to reduce errors between experimental
outcomes and theoretical results, the purity would be the best choice at first
glance, since it is just obtained by matrix multiplication $\left(  Tr\left[
\rho_{s}^{2}\left(  t\right)  \right]  \right)  $ instead of diagonalization
or maximization processes, which must be performed to calculate the other
physical quantities. If the results obtained from the purity are not decisive,
one should perform an analysis of the other physical quantities to be able to
determine the initial correlated state. Thus, by employing such a procedure
described above, one is able to infer if there is initial
system-environment correlation and which is its nature.

It is interesting to remark what would occur if other kinds of interactions in
the Hamiltonian were considered. For example, an important class of such
Hamiltonians is found when the system is coupled to a dephasing environment,
where such a coupling is described by the Pauli matrix $\sigma_{z}$. This kind
of interaction is very common to model the dynamics of solid states systems,
\emph{e.g.}, quantum dots~\cite{qds} and Josephson junctions \cite{jj}. If we
consider this type of interaction together with the initial conditions treated
in this work, we are not able to identify the initial correlation between the
system and the environment. This shortcoming occurs because the initial
states of Eqs.~(\ref{rhoa})-(\ref{rhod}) are written as an ensemble of
eigenvectors of $\sigma_{z}$, thus no difference on the dynamics will be
observed for this type of interaction. However, if other initial conditions
were considered, \emph{e.g.}, initial states represented by an ensemble of
eigenvectors of $\sigma_{x}$, the same methodology could have been applied.

\section{Conclusion}

In summary, we studied a way to determine the existence of initial correlation
between a two-level atom and a Lorentzian structured reservoir at zero
temperature. We found that the dynamics of quantum and classical correlations
have signatures of the initial system-environment correlations. Moreover, we
showed that it is possible to identify the distinct initial correlated states
by considering a probe qubit interacting with the same reservoir. We verified
that, in our case, the purity might be the best choice to witness the initial
system-environment correlations because this quantity reduces errors when the
experimental outcomes are compared to the theoretical results, although the
purity is not always decisive to determine the nature of the initial
correlations. In such a situation, the dynamics of classical or quantum correlations measures for the bipartite system (two atoms) can be used to
distinguish among different initial states. Finally, as we have only presented
a numerical study, we hope that our work stimulates new researches in this
area, specially in order to obtain analytical results that probes the efficiency of purity and the measures of correlations as witness to initial system-environment correlations for more general states and Hamiltonians. We also observed that both the
purity and the measures of correlations become less efficient when the
interaction between the qubits and the reservoir is weakened, so that it is
necessary to find a more precise witness to the initial qubit-reservoir
correlation in the weak coupling regime.

This work is supported by FAPESP and CNPq through the National Institute for
Science and Technology of Quantum Information (INCT-IQ).

\end{document}